%% file: main.tex
\documentclass[10pt,twocolumn,letterpaper]{article}

\usepackage{iccv}
\usepackage{times}
\usepackage{epsfig}
\usepackage{graphicx}
\usepackage{amsmath}
\usepackage{amssymb}

\usepackage{cite}
\usepackage{url}
\usepackage{booktabs}
\usepackage{multirow}
\usepackage[table,xcdraw,dvipsnames]{xcolor}
\definecolor{NewColor1}{HTML}{A0A2ED}
\usepackage{placeins}

\usepackage{paralist}
\usepackage[inline]{enumitem}

\usepackage[vlined,linesnumbered,ruled,resetcount]{algorithm2e}

\usepackage{bm}
\usepackage{blindtext}

\usepackage{breqn}

\usepackage{float}
\usepackage[caption = false]{subfig}

\makeatletter
\newcommand{\clearsubcaptcounter}{\setcounter{sub\@captype}{0}}
\makeatother

\usepackage{etoolbox, siunitx}

\usepackage{amssymb}
\usepackage{pifont}
\usepackage[pagebackref=true,breaklinks=true,letterpaper=true,colorlinks,bookmarks=false]{hyperref}

\iccvfinalcopy 

\def\iccvPaperID{26} 

\ificcvfinal\fi

\begin{document}

\title{Estimating Leaf Water Content using Remotely Sensed Hyperspectral Data}

\author{Vishal Vinod\\
Indian Institute of Science Bangalore\\
{\tt\small vishalvinod@acm.org}
\and
Rahul Raj, Rohit Pingale, Adinarayana Jagarlapudi\\
Indian Institute of Technology, Bombay\\
{\tt\small \{rahul\_raj,rohitpingale103,adi\}@iitb.ac.in}
}

\maketitle
\ificcvfinal\thispagestyle{empty}\fi


\section{Introduction}
\label{sec:intro}
    \input{Sections/1_Intro}

\section{Related Work}
\label{sec:related}
    \input{Sections/2_Related}

\section{Methodology}
\label{sec:method}
    \input{Sections/4_Method}

\vspace{-2mm}

\section{Conclusion}
\label{sec:conclusion}
    \input{Sections/6_Conclusion}

{\small
\bibliographystyle{ieee_fullname}
\bibliography{egbib}
}

\end{document}

%% file: Sections/1_Intro.tex
Plant water stress may occur due to the limited availability of water to the roots/soil or due to increased transpiration. These factors adversely affect plant physiology and photosynthetic ability to the extent that it has been shown to have inhibitory effects in both growth and yield \cite{osakabe2014response}. Early identification of plant water stress status enables suitable corrective measures to be applied to obtain the expected crop yield. Further, improving crop yield through precision agriculture methods is a key component of climate policy and the UN sustainable development goals \cite{UNSDG2030}. Leaf water content (LWC) is a measure that can be used to estimate water content and identify stressed plants. LWC during the early crop growth stages is an important indicator of plant productivity and yield. The effect of water stress can be instantaneous \cite{ma2018sequence}, affecting gaseous exchange or long-term, significantly reducing yield \cite{reddy2003physiological, hsiao1973plant, osakabe2014response}. It is thus necessary to identify potential plant water stress during the early stages of growth \cite{ma2018sequence} to introduce corrective irrigation and alleviate stress. LWC is also useful for identifying plant genotypes that are tolerant to water stress and salinity by measuring the stability of LWC even under artificially induced water stress \cite{osakabe2014response, sakamoto2004arabidopsis}. Such experiments generally employ destructive procedures to obtain the LWC, which is time-consuming and labor intensive. Accordingly, this research has developed a non-destructive methodology to estimate LWC from UAV-based hyperspectral data.

%% file: Sections/2_Related.tex
Previous works have explored water sensitive wavelengths \cite{jin2017determination, moshou2014water} and indices \cite{kim2015highly} for estimating LWC. However, most of the work is done with 400-2500 nm data. These sensors are expensive and arduous to maintain. \cite{lowe2017hyperspectral} discuss the differences between greenhouse and real-world outdoor monitoring with UAVs, and show drought prediction from spectral band data and general vegetative indices (NDVI, PRI, etc.). Recent works \cite{ishida2018novel, vinod2021mining} present methods for classification using Hyperspectral images (HSI). \cite{moharana2016spatial} shows that the spatial variability of chlorophyll from HSI may correlate with LWC. Several works \cite{vigneau2011potential, yilmaz2008remote, zygielbaum2009non} describe methods to estimate water and nitrogen levels in the crop using non-destructive means. However, very few of them have used cross-sensor data modeling. Accordingly, this work investigates water-sensitive indices from hand-held spectroradiometer data that affect water content and train ensemble regressors to compute LWC. A threshold of $79\%$ \cite{ma2018sequence} is used to classify the water stress status of each pixel in the UAV-captured hyperspectral image based on the LWC estimation for the six-leaf growth stage. Further, this work also presents qualitative results for temporal variation of LWC during various stages of crop growth for water stress forecasting. This is the first work to estimate LWC by identifying pure-pixel water-sensitive indices in the 400-1000nm range of the spectroradiometer data to train ensemble methods that generalize for UAV-captured hyperspectral images in real-world outdoor environments.

%% file: Sections/4_Method.tex
\begin{figure*}[ht]
\begin{center}
\centerline{\includegraphics[height=5cm]{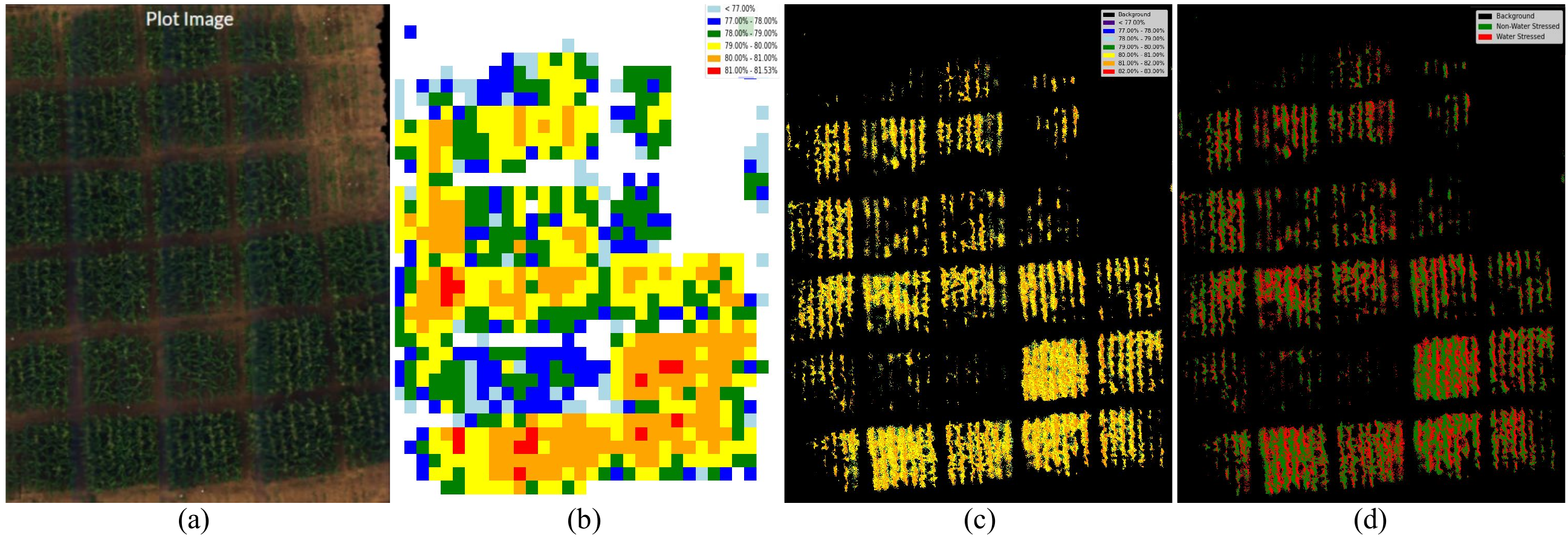}}
\caption{Pixel-wise Crop Water Stress Prediction for 20 Nov 2018 (DAS=36) data. (a) Aerial image of plot, (b) Super-pixel $1m \times 1m$ plot computed using Raj \etal \cite{raj2021leaf2} method used as pseudo ground-truth for LWC estimation, (c) The LWC estimation plot with an NDVI threshold of $0.7$ for the proposed method using GBM, (d) Water stress classification using 79\% threshold.}
\label{fig:LWC}
\end{center}
\vskip -0.3in
\end{figure*}


\textbf{Dataset}. \label{sec:dataset} The experimental farmland consists of 27 specialized treatment plots with varying levels of fertilizer and water provided at regular intervals. The data collected includes hyperspectral data recorded from a hand-held spectroradiometer and a UAV-based Hyperspectral camera to capture HSI of the entire farmland at various stages of growth. The spectroradiometer data consists of 483 data points each with 382 Hyperspectral band data, DAS (Days After Sowing), Nitrogen, Carbon, CNRatio and LWC. The LWC ground-truth value is obtained from destructive distillation of the leaves from the respective plots.

\vskip 0.025in

\textbf{Water-sensitive Indices}.
\label{indices}
Water absorption wavelengths in the 400-1000nm range are more practical to collect due to minimal overhead in maintenance and cost of the sensors. Previous works \cite{raj2021leaf,thenkabail2002evaluation} show that there is a secondary set of water-sensitive wavelengths in the 400-1000nm range. Thus, this work uses the s range for data collection and conduct 400-1000nm range correlation analysis to identify water-sensitive indices. The indices used are tabulated in Table \ref{tab:indices}. NDVI has been shown to correlate with water stress \cite{caturegli2020effects,ballester2019monitoring,genc2011determination} and is also a common indicator of crop health \cite{agapiou2012evaluation}. The differences in water treatment in the 27 specialized treatment plots may lead to variations in foliage color, and under these conditions the green and red normalized indices \cite{amatya2012hyperspectral} are more water-sensitive than NDVI \cite{ballester2019monitoring,katsoulas2016crop}, hence the Green NDVI and Red Edge \cite{zhang2019estimation} indices have been used in this work. There exist secondary water-sensitive bands at 950-970nm \cite{amatya2012hyperspectral,stomp2007colorful,genc2011determination} and several works \cite{raj2021leaf,katsoulas2016crop,thenkabail2002evaluation,amatya2012hyperspectral} show the direct decrease in Water Index (WI) with water stress. The Renormalized DVI indices (RDVI) \cite{roujean1995estimating} and Optimized Soil Adjusted Vegetation Index (OSAVI) \cite{rondeaux1996optimization,raj2020precision} have shown to be good predictors of LWC. Further, rigorous analytic experiments show a strong correlation (coefficient: 0.84) among MTVI2, OSAVI and RDVI. This work builds on the findings in \cite{raj2021leaf2} to identify NIR1, RedBlue and WaterBand indices to have a strong correlation with LWC.

\begin{table}[t]
\caption{The indices used to estimate leaf water content.}
\label{tab:indices}
\vskip 0.15in
\begin{center}
\begin{small}
\begin{sc}
\scalebox{0.99}{
\begin{tabular}{@{}c|c@{}}
\toprule
\textbf{Indices} & \textbf{Formula} \\
\toprule \toprule
NDVI    & $(\rho_{\textrm{nir}}-\rho_{\textrm{red}})/(\rho_{\textrm{nir}}+\rho_{\textrm{red}})$ \\
\midrule
Green NDVI & $(\rho_{\textrm{nir}}-\rho_{\textrm{green}})/(\rho_{\textrm{nir}}+\rho_{\textrm{green}})$ \\
\midrule
RDVI & $(\rho_{\textrm{nir}}-\rho_{\textrm{red}})/(\rho_{\textrm{nir}}+\rho_{\textrm{red}})^{1/2}$ \\
\midrule 
MTVI2 & $\frac{1.5\times(1.2\times(\rho_{\textrm{nir}}-\rho_{\textrm{green}})-2.5\times(\rho_{red}-\rho_{green})]}{[(2\times\rho_{\textrm{nir}}+1)^2-(6\times\rho_{\textrm{nir}}-5\times\rho_{red}^{1/2})-0.5]^{1/2}} $ \\
\midrule
 Water Index & $\rho_{\textrm{900}}/\rho_{\textrm{970}}$ \\
\midrule
NPCI & $(\rho_{\textrm{680}}-\rho_{\textrm{430}})/(\rho_{\textrm{680}}+\rho_{\textrm{430}})$ \\
\midrule
OSAVI & $[1.16* (\rho_{\textrm{800}} - \rho_{\textrm{670}})]/(\rho_{\textrm{800}} + \rho_{\textrm{670}} + 0.16)$ \\
\midrule
Red Edge & $\rho_{\textrm{750}}/\rho_{\textrm{710}}$ \\
\midrule
NIR1 & $(\rho_{\textrm{800}} - \rho_{\textrm{847}})/(\rho_{\textrm{800}} + \rho_{\textrm{847}})$ \\
\midrule
RedBlue & $(\rho_{\textrm{660}} - \rho_{\textrm{420}})/(\rho_{\textrm{660}} + \rho_{\textrm{420}})$ \\
\midrule
Water Band & $(\rho_{\textrm{791}} - \rho_{\textrm{970}})/(\rho_{\textrm{791}} + \rho_{\textrm{970}})$ \\
\bottomrule
\end{tabular}
}
\end{sc}
\end{small}
\end{center}
\vskip -0.25in
\end{table}

\textbf{Estimating Leaf Water Content}.
\label{lwc}
\label{lwc-status}
Ensemble methods and regressors were trained on the handheld spectroradiometer data (see Table \ref{tab:lwc}). Note that the experiment with three indices corresponds to using NIR1, RedBlue and WaterBand indices, and eight indices correspond to the remaining indices listed in Table \ref{tab:indices}. The Gradient Boosting \cite{friedman2001elements,friedman2001greedy} method demonstrated the best RMSE and $R^2$ metrics for the eleven indices experiment (see Table \ref{tab:lwc}). The trained Gradient Boost method is used for the inference stage on the geo-tagged UAV-captured high-resolution hyperspectral images (HSI). The performance is compared with the super-pixel 1m x 1m plot computed by Raj \etal \cite{raj2021leaf2} as psuedo ground truth for the surrounding pixels. As discussed in Sec.(\ref{sec:related}), the threshold of $79\%$ water content during the six-leaf stage DAS=36 (DAS: Days After Sowing) of the crop to classify the water stress status. Fig.(\ref{fig:LWC}) shows LWC inference and stress prediction on the UAV-captured hyperspectral images for DAS=36 compared with the $1m\times1m$ super-pixel plot. 




\begin{table}[t]
\vskip -0.05in
\begin{center}
\begin{small}
\begin{sc}
\scalebox{0.82}{
\begin{tabular}{c|cc|cc|cc}
\toprule
Regression & \multicolumn{2}{c}{8 Indices} & \multicolumn{2}{c}{3 Indices} & \multicolumn{2}{c}{11 Indices} \\
Algorithm & RMSE & $R^2$ & RMSE & $R^2$ & RMSE & $R^2$\\
\midrule
Gradient Boost & 0.0344 & 0.641 & \textbf{0.0158} & \textbf{0.925} & \textbf{0.0153} & \textbf{0.930}\\ 
Lasso Cross-Val & 0.0361 & 0.610 & 0.0172 & 0.911 & 0.0171 & 0.912\\ 
Random Forest & 0.0354 & 0.622 & 0.0165 & 0.918 & 0.0161 & 0.922\\ 
Stacked Methods & \textbf{0.0338} & \textbf{0.658} & 0.0162 & 0.921 & 0.0154 & 0.929\\ 

\bottomrule
\end{tabular}
}
\end{sc}
\end{small}
\end{center}
\vskip -0.12in
\caption{Comparing LWC estimation prediction with different subset of indices. We report the corresponding RMSE and $R^2$ metrics.}
\label{tab:lwc}
\vskip -0.15in
\end{table}

%% file: Sections/6_Conclusion.tex
This work identifies indices that are correlated with leaf water content and water stress. First, the values of the indices from the spectral bands of the hand-held spectroradiometer data are computed. Then, ensemble methods are trained on this dataset with high-quality LWC ground truth values and the $1m \times 1m$ super-pixel pseudo ground truth verifying the improvement in RMSE and $R^2$ metric using the proposed set of eleven indices. The trained model is evaluated on UAV-captured hyperspectral images for varying stages of growth and DAS to compute pixel-level LWC statistics for each treatment plot. The comparison between the mean predictions on the HSI with the psuedo ground truth mean water content values for each plot verify the performance of the method in estimating LWC effectively. 


%% file: main.bbl
\begin{thebibliography}{10}\itemsep=-1pt

\bibitem{UNSDG2030}
United nations sustainable development goals: Food security and nutrition and
  sustainable agriculture.
\newblock https://sustainabledevelopment.un.org/topics/foodagriculture.
\newblock Online; Accessed: 2021-05-30.

\bibitem{agapiou2012evaluation}
Athos Agapiou, Diofantos~G Hadjimitsis, and Dimitrios~D Alexakis.
\newblock Evaluation of broadband and narrowband vegetation indices for the
  identification of archaeological crop marks.
\newblock {\em Remote sensing}, 4(12):3892--3919, 2012.

\bibitem{amatya2012hyperspectral}
Suraj Amatya, Manoj Karkee, Ashok~K Alva, Peter Larbi, and Bikram Adhikari.
\newblock Hyperspectral imaging for detecting water stress in potatoes.
\newblock In {\em 2012 Dallas, Texas, July 29-August 1, 2012}, page~1. American
  Society of Agricultural and Biological Engineers, 2012.

\bibitem{ballester2019monitoring}
Carlos Ballester, James Brinkhoff, Wendy~C Quayle, and John Hornbuckle.
\newblock Monitoring the effects of water stress in cotton using the green red
  vegetation index and red edge ratio.
\newblock {\em Remote Sensing}, 11(7):873, 2019.

\bibitem{caturegli2020effects}
Lisa Caturegli, Stefania Matteoli, Monica Gaetani, Nicola Grossi, Simone Magni,
  Alberto Minelli, Giovanni Corsini, Damiano Remorini, and Marco Volterrani.
\newblock Effects of water stress on spectral reflectance of bermudagrass.
\newblock {\em Scientific Reports}, 10(1):1--12, 2020.

\bibitem{friedman2001elements}
Jerome Friedman, Trevor Hastie, Robert Tibshirani, et~al.
\newblock {\em The elements of statistical learning}, volume~1.
\newblock Springer series in statistics New York, 2001.

\bibitem{friedman2001greedy}
Jerome~H Friedman.
\newblock Greedy function approximation: a gradient boosting machine.
\newblock {\em Annals of statistics}, pages 1189--1232, 2001.

\bibitem{genc2011determination}
Levent Genc, Kursad Demirel, Gokhan Camoglu, Serafettin Asik, Scot Smith,
  et~al.
\newblock Determination of plant water stress using spectral reflectance
  measurements in watermelon (citrullus vulgaris).
\newblock {\em American-Eurasian Journal of Agricultural \& Environmental
  Sciences}, 11(2):296--304, 2011.

\bibitem{hsiao1973plant}
Theodore~C Hsiao.
\newblock Plant responses to water stress.
\newblock {\em Annual review of plant physiology}, 24(1):519--570, 1973.

\bibitem{ishida2018novel}
Tetsuro Ishida, Junichi Kurihara, Fra~Angelico Viray, Shielo~Baes Namuco,
  Enrico~C Paringit, Gay~Jane Perez, Yukihiro Takahashi, and Joel~Joseph
  Marciano~Jr.
\newblock A novel approach for vegetation classification using uav-based
  hyperspectral imaging.
\newblock {\em Computers and electronics in agriculture}, 144:80--85, 2018.

\bibitem{jin2017determination}
Xiaoli Jin, Chunhai Shi, Chang~Yeon Yu, Toshihiko Yamada, and Erik~J Sacks.
\newblock Determination of leaf water content by visible and near-infrared
  spectrometry and multivariate calibration in miscanthus.
\newblock {\em Frontiers in plant science}, 8:721, 2017.

\bibitem{katsoulas2016crop}
Nikolaos Katsoulas, Angeliki Elvanidi, Konstantinos~P Ferentinos, Murat Kacira,
  Thomas Bartzanas, and Constantinos Kittas.
\newblock Crop reflectance monitoring as a tool for water stress detection in
  greenhouses: A review.
\newblock {\em biosystems engineering}, 151:374--398, 2016.

\bibitem{kim2015highly}
David~M Kim, Hairong Zhang, Haiying Zhou, Tommy Du, Qian Wu, Todd~C Mockler,
  and Mikhail~Y Berezin.
\newblock Highly sensitive image-derived indices of water-stressed plants using
  hyperspectral imaging in swir and histogram analysis.
\newblock {\em Scientific reports}, 5(1):1--11, 2015.

\bibitem{lowe2017hyperspectral}
Amy Lowe, Nicola Harrison, and Andrew~P French.
\newblock Hyperspectral image analysis techniques for the detection and
  classification of the early onset of plant disease and stress.
\newblock {\em Plant methods}, 13(1):1--12, 2017.

\bibitem{ma2018sequence}
Xueyan Ma, Qijin He, and Guangsheng Zhou.
\newblock Sequence of changes in maize responding to soil water deficit and
  related critical thresholds.
\newblock {\em Frontiers in plant science}, 9:511, 2018.

\bibitem{moharana2016spatial}
Shreedevi Moharana and Subashisa Dutta.
\newblock Spatial variability of chlorophyll and nitrogen content of rice from
  hyperspectral imagery.
\newblock {\em ISPRS Journal of Photogrammetry and Remote Sensing}, 122:17--29,
  2016.

\bibitem{moshou2014water}
Dimitrios Moshou, Xanthoula-Eirini Pantazi, Dimitrios Kateris, and Ioannis
  Gravalos.
\newblock Water stress detection based on optical multisensor fusion with a
  least squares support vector machine classifier.
\newblock {\em Biosystems Engineering}, 117:15--22, 2014.

\bibitem{osakabe2014response}
Yuriko Osakabe, Keishi Osakabe, Kazuo Shinozaki, and Lam-Son~Phan Tran.
\newblock Response of plants to water stress.
\newblock {\em Frontiers in plant science}, 5:86, 2014.

\bibitem{raj2020precision}
Rahul Raj, Soumyashree Kar, Rohit Nandan, and Adinarayana Jagarlapudi.
\newblock Precision agriculture and unmanned aerial vehicles (uavs).
\newblock In {\em Unmanned Aerial Vehicle: Applications in Agriculture and
  Environment}, pages 7--23. Springer, 2020.

\bibitem{raj2021leaf2}
Rahul Raj, Jeffrey~P Walker, Rohit Pingale, Rohit Nandan, Balaji Naik, and
  Adinarayana Jagarlapudi.
\newblock Leaf area index estimation using top-of-canopy airborne rgb images.
\newblock {\em International Journal of Applied Earth Observation and
  Geoinformation}, 96:102282, 2021.

\bibitem{raj2021leaf}
Rahul Raj, Jeffrey~P Walker, Vishal Vinod, Rohit Pingale, Balaji Naik, and
  Adinarayana Jagarlapudi.
\newblock Leaf water content estimation using top-of-canopy airborne
  hyperspectral data.
\newblock {\em International Journal of Applied Earth Observation and
  Geoinformation}, 102:102393, 2021.

\bibitem{reddy2003physiological}
TY Reddy, VR Reddy, and V Anbumozhi.
\newblock Physiological responses of groundnut (arachis hypogea l.) to drought
  stress and its amelioration: a critical review.
\newblock {\em Plant growth regulation}, 41(1):75--88, 2003.

\bibitem{rondeaux1996optimization}
Genevi{\`e}ve Rondeaux, Michael Steven, and Fr{\'e}d{\'e}ric Baret.
\newblock Optimization of soil-adjusted vegetation indices.
\newblock {\em Remote sensing of environment}, 55(2):95--107, 1996.

\bibitem{roujean1995estimating}
Jean-Louis Roujean and Francois-Marie Breon.
\newblock Estimating par absorbed by vegetation from bidirectional reflectance
  measurements.
\newblock {\em Remote sensing of Environment}, 51(3):375--384, 1995.

\bibitem{sakamoto2004arabidopsis}
Hideki Sakamoto, Kyonoshin Maruyama, Yoh Sakuma, Tetsuo Meshi, Masaki Iwabuchi,
  Kazuo Shinozaki, and Kazuko Yamaguchi-Shinozaki.
\newblock Arabidopsis cys2/his2-type zinc-finger proteins function as
  transcription repressors under drought, cold, and high-salinity stress
  conditions.
\newblock {\em Plant physiology}, 136(1):2734--2746, 2004.

\bibitem{stomp2007colorful}
Maayke Stomp, Jef Huisman, Lucas~J Stal, and Hans~CP Matthijs.
\newblock Colorful niches of phototrophic microorganisms shaped by vibrations
  of the water molecule.
\newblock {\em The ISME journal}, 1(4):271--282, 2007.

\bibitem{thenkabail2002evaluation}
Prasad~S Thenkabail, Ronald~B Smith, and Eddy De~Pauw.
\newblock Evaluation of narrowband and broadband vegetation indices for
  determining optimal hyperspectral wavebands for agricultural crop
  characterization.
\newblock {\em Photogrammetric engineering and remote sensing}, 68(6):607--622,
  2002.

\bibitem{vigneau2011potential}
Nathalie Vigneau, Martin Ecarnot, Gilles Rabatel, and Pierre Roumet.
\newblock Potential of field hyperspectral imaging as a non destructive method
  to assess leaf nitrogen content in wheat.
\newblock {\em Field Crops Research}, 122(1):25--31, 2011.

\bibitem{vinod2021mining}
Vishal Vinod, Vipul Gaurav, Tushar Sharma, and Savita Choudhary.
\newblock Mining intelligent patterns using svac for precision agriculture and
  optimizing irrigation (student abstract).
\newblock In {\em Proceedings of the AAAI Conference on Artificial
  Intelligence}, volume~35, pages 15909--15910, 2021.

\bibitem{yilmaz2008remote}
M~Tugrul Yilmaz, E~Raymond Hunt~Jr, and Thomas~J Jackson.
\newblock Remote sensing of vegetation water content from equivalent water
  thickness using satellite imagery.
\newblock {\em Remote Sensing of Environment}, 112(5):2514--2522, 2008.

\bibitem{zhang2019estimation}
F Zhang and G Zhou.
\newblock Estimation of vegetation water content using hyperspectral vegetation
  indices: A comparison of crop water indicators in response to water stress
  treatments for summer maize.
\newblock {\em BMC ecology}, 19(1):18, 2019.

\bibitem{zygielbaum2009non}
Arthur~I Zygielbaum, Anatoly~A Gitelson, Timothy~J Arkebauer, and Donald~C
  Rundquist.
\newblock Non-destructive detection of water stress and estimation of relative
  water content in maize.
\newblock {\em Geophysical research letters}, 36(12), 2009.

\end{thebibliography}
